\documentclass[9pt,twocolumn,twoside]{pnas-new}

\articletype{Physical Sciences | Engineering}
\templatetype{pnasresearcharticle}

\usepackage{amsmath,amssymb}
\usepackage{booktabs}
\usepackage{tabularx}
\usepackage{float}
\usepackage[section]{placeins}
\usepackage{tikz}
\usepackage{siunitx}
\usetikzlibrary{arrows.meta,positioning,shapes.multipart}
\setboolean{displaywatermark}{false}
\begin{document}

\title{Mycoponically Integrated Network Device for Multimodal Biosensing using Electrophysiological Mycelial Networks}

\author[a]{Zihan Oliver Zeng}
\author[b]{D. Marshall Porterfield}
\author[b,1]{Upinder Kaur}

\affil[a]{Mechanical Engineering, Purdue University, West Lafayette, IN 47906, USA}
\affil[b]{Agricultural and Biological Engineering, Purdue University, West Lafayette, IN 47906, USA}

\leadauthor{Kaur}

\significancestatement{Living fungal mycelium responds to chemical, optical, mechanical, thermal, and biological stimuli through a shared membrane-voltage mechanism. However, its use as a biosensor has been constrained by the absence of an interface that simultaneously sustains metabolism, standardizes electrode coupling, and tolerates mechanical damage. We engineered the Mycoponically Integrated Network Device (MIND), using porous ceramics as an antimicrobial liquid nutrient delivery platform with integrated electrophysiology, converting the mycelium into a multimodal, self-repairing, multi-domain biosensor. MIND distinguishes 14 stimuli, with calibration across five fungal species/clades. We also show function recovery within 72 h of mechanical damage and sustained operation beyond 11 months. MIND lays the foundation for transforming mycelial networks into multi-domain biosensors.}

\authorcontributions{Z.O.Z. collected data, analyzed data, and helped in writing the manuscript; D.M.P. contributed to the design, conceptualization, and the biological interpretation of the data, and edited the draft manuscript; U.K. led the design conceptualization, fabrication, supervised the project, guided engineering design and biological interpretation, and wrote the draft manuscript.}
\authordeclaration{The authors declare that they have registered the associate intellectual property with the Purdue Office of Commercialization, and a provisional patent has been filed.}
\correspondingauthor{\textsuperscript{1}To whom correspondence should be addressed. E-mail: porterf@purdue.edu, kauru@purdue.edu}


\begin{abstract}
Living mycelial filaments integrate chemical, optical, mechanical, thermal, and biological information via electrophysiological cellular trans-membrane potential. The challenge is to create a mycology interface that sustains metabolism, standardizes electrode geometry, and tolerates mechanical damage. Using mycoponics we overcome these factors that limited prior demonstrations to single modalities, and operational windows of days to weeks. We present MIND, an engineered biophysical interface integrating antimicrobial nutrient delivery (ceramic size exclusion) with non-invasive electrophysiology, in cylindrical (MINDTube) and planar (MINDPixel) form-factors. The platform sustains colonized \textit{Pleurotus ostreatus} mycelium beyond 11 months and distinguishes 14 stimulus classes from a single unmodified device. Steady-state intensity responses follow Hill-type calibration functions across five phylogenetically diverse fungi grown on the identical interface, making strain selection a tunable design parameter. Multichannel decoding from the standardized electrode geometry recovers stimulus duration, location, and trajectory. Continuous nutrition provided by mycoponics recovered complete electrophysiological function within 72 h after mechanical excision. MIND converts living mycelium networks into universal, self-repairing, biosensors.
\end{abstract}

\dates{This manuscript was compiled on \today}
\doi{\url{www.pnas.org/cgi/doi/10.1073/pnas.XXXXXXXXXX}}

\maketitle
\thispagestyle{firststyle}
\ifthenelse{\boolean{shortarticle}}{\ifthenelse{\boolean{singlecolumn}}{\abscontentformatted}{\abscontent}}{}

\Firstpage

Conventional sensors are engineered for defined targets. Monitoring chemical, optical, mechanical, and thermal domains often requires a suite of distinct transducers. This design is effective when the target is known, but it cannot resolve unknown stimuli without prior selection of the appropriate detector. A multi-domain sensor must first identify what is present before estimating how much is present. Fungal mycelium provides a biological path toward this capability as different stimuli evoke distinct electrical response patterns. Metabolically active hyphae form an electrically coupled lattice in which local ionic perturbations propagate as distributed extracellular voltage transients~\cite{Buffi2025ElectricalSignalingFungi, Adamatzky2022FungalElectronics}. Photoreceptors, ion channels, chemosensory proteins, and mechanosensors each couple into the same membrane-voltage transduction pathway, often involving calcium channels~\cite{Qi2020PoWC, Corrochano2019LightFungalWorld, Vu2015Cch1Mid1, Zhao2024FungalPma1}. This organization allows a single mycelial network to integrate cellular control across multiple physical and biological domains.

The biological convergence has motivated the use of living mycelia as electrophysiological sensors. Empirical work has shown that living mycelia generate measurable electrical responses to the physical and chemical stimuli and can support biohybrid actuation~\cite{Mishra2024MyceliumBiohybridRobots, GandiaAdamatzky2024FungalSkin, Phillips2023MoistureResponse}. However, most studies report these signals as qualitative correlations between an applied stimulus and a voltage pattern. Decoding complex information such as stimulus location, intensity, duration, or simultaneous inputs, from the bioelectrical signal remains unresolved~\cite{Adamatzky2022FungalElectronics, Buffi2025ExternalStimuli, Buffi2025ElectricalSignalingFungi}. Quantitative applications in sensing, robotics, and neural-inspired computation require a robust, integrative architecture that sustains metabolic activity, standardizes electrode coupling, and produces signals sufficiently stable for calibration.

\begin{figure*}[t]
    \centering
    \includegraphics[trim={40 35 20 20},clip,width=\linewidth]{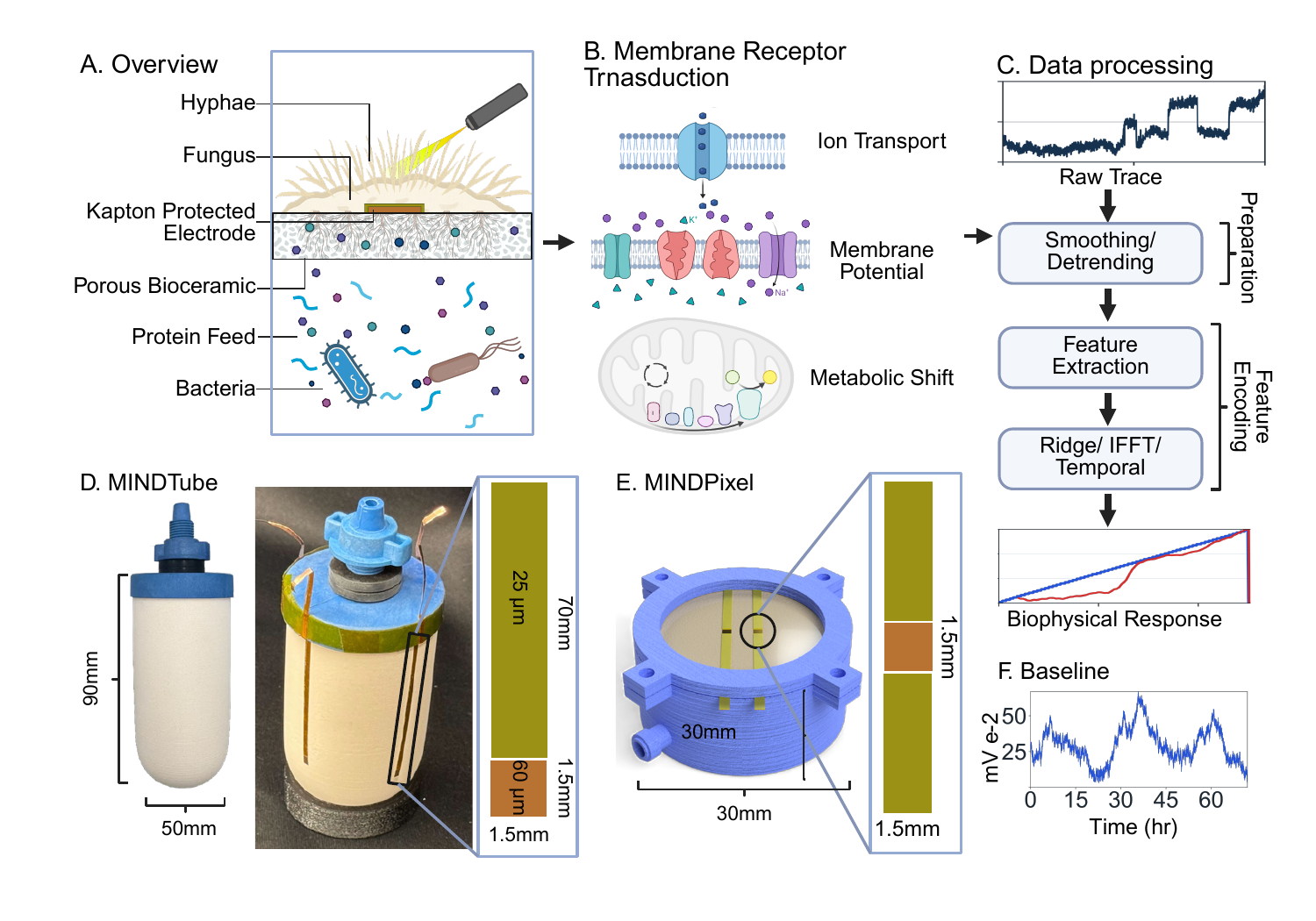}
    \caption{Overview of the MIND platform and analysis workflow.
    (\textit{A}) Schematic of the living mycelial sensing interface and passive extracellular readout architecture; the bioceramic substrate builds on prior mycoponics work showing bacterial filtration capability~\cite{Sanchez2026Mycoponics}.
    (\textit{B}) Conceptual stimulus-to-signal pathway linking environmental input to extracellular electrical response.
    (\textit{C}) Analysis workflow from raw acquisition through preprocessing, feature extraction, decoding, and inferred output.
    (\textit{D}) MINDTube morphology, representative implementation, and key dimensions.
    (\textit{E}) MINDPixel morphology, representative implementation, and key dimensions.}
    \label{fig:overview_workflow}
\end{figure*}

We present \textbf{MIND} (Mycoponically Integrated Network Device), a dynamic multichannel extracellular recording platform that addresses fungal sensing\Endparasplit requirements through an integrated growth and recording architecture (see Fig.~\ref{fig:overview_workflow}). A porous bioceramic substrate delivers nutrients to the mycelial layer by gravity-fed capillary perfusion, supporting sustained metabolic activity without substrate replacement\cite{Sanchez2026Mycoponics}. Stable electrode contact with the colonized surface enables differential extracellular recordings over long durations. Two architectures implement this principle. MINDTube (Fig.~\ref{fig:overview_workflow}D) supports continuous mycelial growth on a cylindrical bioceramic substrate with four differential circumferential electrode channels. MINDPixel (Fig.~\ref{fig:overview_workflow}E) uses flat bioceramic disks with discrete electrically isolated mycelial pixels and localized electrode pairs. These systems sustained colonized \textit{Pleurotus ostreatus} mycelial networks with extracellular recordings for more than 11 months at the time of this submission. Using MINDTube, we characterize bioelectrical responses to 14 input stimuli spanning chemical, optical, mechanical, thermal, and biological domains. We replicate the bioelectrical responses for four additional fungal species from different phylogenetic clades. Using MINDPixel, we calibrate morphology-specific intensity-response relationships and decode stimulus location, duration, and continuous trajectory from multichannel bioelectrical features. We further demonstrate self-repair after mechanical damage, linking long-term viability to functional recovery in living hybrid bioelectric devices.

\section*{Results}\label{sec:results}

\subsection*{MIND provides stable passive readout across fourteen stimulus classes}

We first tested whether recorded voltage deflections originated from living mycelium rather than the bioceramic substrate, stimulus-delivery protocol, or biological sample handling. We then quantified resting baseline behavior, responses across 14 stimulus classes, and recovery after repeated stimulation. Unless otherwise noted, all experiments used blue oyster mycelium (\textit{Pleurotus ostreatus}).


We sterilized bioceramic tubes without mycelium and exposed them to all fourteen stimulus classes under an acquisition protocol identical to that used in all biological experiments. No stimulus condition produced voltage deflections above the noise floor. For the biological stimulus control, we treated harvested maize leaves (\textit{Zea mays}) with isopropyl alcohol and dried them for 24 hours before placement adjacent to the sensor. These treated leaves also produced no measurable response. Together, these controls indicate that the reported voltage deflections arose from the living mycelial layer.

With the mycelial network present, MIND produced a resting potential that fluctuated around a slowly varying mean (Fig.~\ref{fig:overview_workflow}F). Median subtraction centered this baseline before downstream analysis. The corrected signal then provided the reference for quantifying stimulus-evoked voltage deflections. Its properties reflect biophysical differences in electrophysiological activity across fungal species.

\begin{figure*}[t!]
    \centering
    \includegraphics[trim={325 160 325 160},clip,width=.8\linewidth]{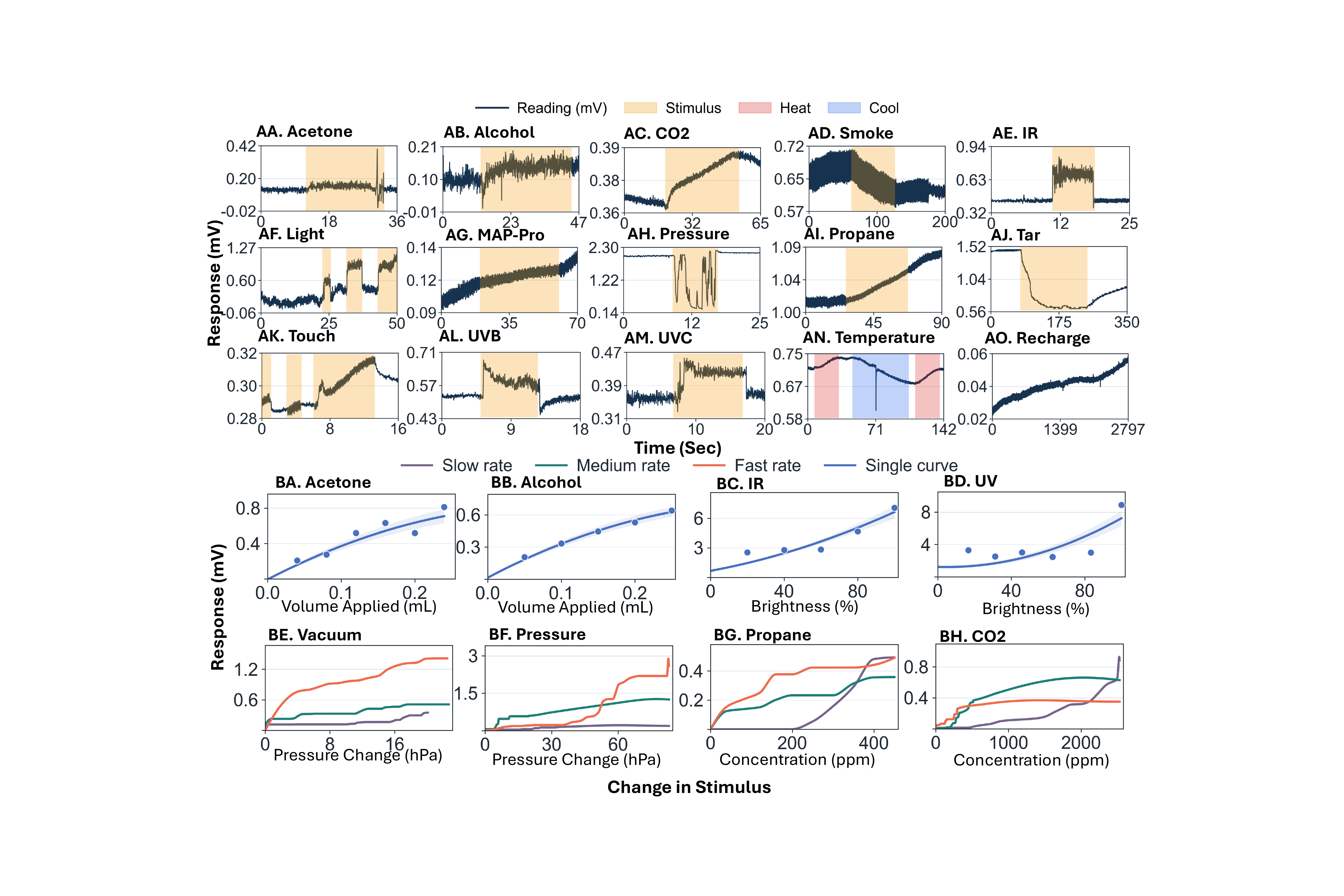}
    \caption{Representative bioelectrical responses across stimulus classes and representative stimulus dose--response assays. Experiments were conducted with MINDTube modules supporting \textit{Pleurotus ostreatus} (Blue Oyster). (\textit{A}) Representative single-stimulus bioelectrical traces for 14 stimulus classes and the recharge case; shaded regions mark the stimulus-on interval. (\textit{B}) Representative response-amplitude sweeps for selected modalities, showing bioelectrical output as a function of applied stimulus level. For modalities tested at multiple loading rates, slow/medium/fast rates were pressure 0.5, 2.0, 4.3\,hPa\,s$^{-1}$; vacuum 0.16, 0.62, 1.30\,hPa\,s$^{-1}$; CO\textsubscript{2} 11, 74, 188\,ppm\,s$^{-1}$; and propane 5, 20, 60\,ppm\,s$^{-1}$.}
    \label{fig:stimuli_overview}
\end{figure*}

Fourteen stimuli spanning five domains were applied to MINDTube without hardware modifications, with 30 independent events for each stimulus. Chemical stimuli included acetone ($CH_3COCH_3$), isopropyl alcohol ($(CH_3)_2CHOH$), carbon dioxide ($CO_2$), propane ($C_3H_8$), smoke, and MAP-Pro\textsuperscript{\textregistered} gas (Worthington Industries, Inc.; 99.5\% propylene, $C_3H_6$; 0.5\% propane, $C_3H_8$)). Optical stimuli included white visible light, near-infrared, ultraviolet B (UVB), and ultraviolet C (UVC) illumination. Mechanical stimuli included pressure and touch. Thermal stimulation was applied by temperature variation. Tar spot (\textit{Phyllachora maydis}) spores provided the biological stimuli. Representative single-stimulus responses are shown in Fig.~\ref{fig:stimuli_overview}A. Modality-specific stimulus--response sweeps are illustrated in Fig.~\ref{fig:stimuli_overview}B.

MIND produced measurable voltage deflections to each stimulus class, with sufficient reproducibility across 30 independent events per condition to resolve systematic differences in amplitude and recovery kinetics. The response hierarchy was nonuniform (Fig.~\ref{fig:stimuli_overview}; Table~\ref{tab:stim-metrics}). Pressure produced the largest deflection, with $\Delta V_{\max} = 2.0 \pm 1.3$\,mV and $\mathrm{SNR}_{\mathrm{mean}} = 15 \pm 11$. Tar spot disease produced the second-largest amplitude ($\Delta V_{\max} = 0.93 \pm 0.15$\,mV) and the slowest recovery ($\tau_{1/e} = 81 \pm 55$\,s). Recovery time constants spanned more than four orders of magnitude across the full stimulus set. Peak amplitude, mean signal-to-noise ratio, and $\tau_{1/e}$ for each stimulus are summarized in Table~\ref{tab:stim-metrics}. Response amplitude and recovery kinetics differed systematically between stimulus classes. Both features carry discriminative information for downstream inference.

MIND resolves how a stimulus is applied, not only its magnitude. Pressure, vacuum, and carbon dioxide were each delivered at three loading rates (pressure: 0.5, 2.0, 4.3\,hPa\,s$^{-1}$; vacuum: 0.16, 0.62, 1.30\,hPa\,s$^{-1}$; CO$_2$: 11, 74, 188\,ppm\,s$^{-1}$) and produced three distinguishable response families each, indicating that MIND encodes stimulus rate as well as magnitude through differential electrode coupling to the rate-dependent receptor responses of the mycelium. Pressure spanned 10.7--82.9\,hPa and reached end-of-range responses of $0.19 \pm 0.01$, $1.25 \pm 0.20$, and $2.59 \pm 0.21$\,mV across the slow, medium, and fast rate curves. Vacuum spanned 4.2--22.8\,hPa and reached $0.35 \pm 0.03$, $0.51 \pm 0.05$, and $1.41 \pm 0.12$\,mV. Carbon dioxide showed the same rate-dependent structure over a proxy range of 0.32--5.06\,hPa, with end-of-range responses of $0.35 \pm 0.02$, $0.63 \pm 0.03$, and $0.88 \pm 0.06$\,mV. IR and isopropyl alcohol did not admit rate subdivision and were treated as single-curve ranked-application stimuli. IR increased across five ordered levels from $2.5 \pm 0.17$ to $7.1 \pm 0.46$\,mV. Isopropyl alcohol increased across five ranked applications (50--250\,$\mu$L) with peak responses from $0.21 \pm 0.016$ to $0.64 \pm 0.043$\,mV. Both curves showed a concave saturating response, consistent with receptor saturation shared across stimuli. The contrast between rate-dependent mechanical and gas responses and receptor-saturating optical and chemical responses is consistent with distinct upstream receptor populations converging on shared electrophysiological machinery.

Continuous nutrient perfusion through the bioceramic enabled the mycelial network to recover ionic balance between stimulation events. After repeated stimulation, the uncorrected baseline exhibited a slow upward drift consistent with recovery of the living interface (Fig.~\ref{fig:stimuli_overview}A). A linear model described this behavior well,
\begin{equation}
\widehat{V}_{\mathrm{base}}(t) = 3.17\times 10^{-3} + 7.93\times 10^{-7}\,t \quad \text{(mV)},
\end{equation}
where $t$ is time in seconds ($R^2 = 0.947$). The fitted slope provides an empirical measure of long-timescale baseline recovery between stimulation events.

\begin{table}[t]
\centering
\caption{MIND Stimulus Response Characteristics}
\textit{}\label{tab:stim-metrics}
\begin{tabular}{l r@{$\,\pm\,$}l r@{$\,\pm\,$}l r@{$\,\pm\,$}l}
\toprule
Stimulus
  & \multicolumn{2}{c}{$\Delta V_{\max}$ [mV]}
  & \multicolumn{2}{c}{$\mathrm{SNR}_{\mathrm{mean}}$}
  & \multicolumn{2}{c}{$\tau_{1/e}$ [s]} \\
\midrule
\multicolumn{7}{l}{\textit{Chemical}} \\
\addlinespace[2pt]
Acetone             & 0.27  & 0.045  & 1.3  & 0.22 & 0.0045 & 0.0038 \\
Isopropyl alcohol   & 0.060 & 0.0075 & 0.85 & 0.21 & 14     & 1.8    \\
CO$_2$              & 0.025 & 0.0050 & 2.4  & 0.43 & 27     & 0.52   \\
Smoke               & 0.15  & 0.072  & 1.03 & 0.49 & 10.3   & 17.3   \\
MAP-Pro gas         & 0.030 & 0.016  & 2.0  & 1.1  & 0.55   & 0.053  \\
Propane             & 0.075 & 0.018  & 1.8  & 0.42 & 40     & 5.3    \\
\addlinespace[2pt]
\multicolumn{7}{l}{\textit{Optical}} \\
\addlinespace[2pt]
IR                  & 0.28  & 0.10   & 4.1  & 1.5  & 0.0058 & 0.00054 \\
White visible light & 0.70  & 0.18   & 2.1  & 1.1  & 6.2    & 0.88    \\
UVB                 & 0.10  & 0.030  & 1.0  & 0.30 & 0.21   & 0.19    \\
UVC                 & 0.80  & 0.011  & 1.9  & 0.26 & 0.0032 & 0.00091 \\
\addlinespace[2pt]
\multicolumn{7}{l}{\textit{Mechanical}} \\
\addlinespace[2pt]
Pressure            & 2.0   & 1.3    & 15   & 11   & 1.8    & 0.23    \\
Touch               & 0.060 & 0.015  & 1.8  & 0.90 & 4.7    & 0.57    \\
\addlinespace[2pt]
\multicolumn{7}{l}{\textit{Thermal}} \\
\addlinespace[2pt]
Temperature         & 0.080 & 0.018  & 1.8  & 0.85 & 39     & 6.4     \\
\addlinespace[2pt]
\multicolumn{7}{l}{\textit{Biological}} \\
\addlinespace[2pt]
Tar spot of maize   & 0.93  & 0.15   & 2.2  & 0.36 & 81     & 55      \\
\bottomrule
\end{tabular}\vspace{-1.0em}
{\raggedright\footnotesize Metrics reported are peak amplitude $\Delta V_{\max}$, mean signal-to-noise ratio $\mathrm{SNR}_{\mathrm{mean}}$, and recovery time constant $\tau_{1/e}$. All entries are mean $\pm$ SD across events.\par}
\end{table}

\subsection*{Intensity responses follow morphology-specific Hill calibration functions}

The fourteen-stimulus survey established that amplitude and recovery kinetics encode stimulus identity. We next tested whether calibrated response amplitude could encode stimulus intensity within a single controlled modality. White visible light served as the model stimulus because intensity, timing, and spatial pattern were independently programmable. The experimental setups for both morphologies are shown in Fig.~\ref{fig:static_setup_calibration_decoding}A,B. All optical experiments in this and the following two subsections were conducted over approximately eight months of continuous platform operation.

Morphology-specific calibration is required for reliable intensity inference. Morphology selection alters the usable operating range, not only physical geometry. We fit the steady-state bioelectrical response of each morphology with a Hill function. Morphology-specific calibration curves are shown in Fig.~\ref{fig:static_setup_calibration_decoding}D,E. For MINDTube, the median fit across 15 independent experiments was
\[
f_{\mathrm{Tube}}(\ell)=\frac{1.22\times\ell^{\,0.94}}{5.40^{\,0.94}+\ell^{\,0.94}},\quad R^2 = 0.9981,
\]
where $\ell$ is light brightness (\% of full LED output). The near-unity exponent ($n = 0.94$) produces a near-linear saturating regime. MINDTube resolves fine intensity differences at low illumination.

For MINDPixel, calibration across six independent modules (18 experiments total) yielded the median fit
\[
f_{\mathrm{Pixel}}(\ell)=\frac{17.60\times\ell^{\,3.90}}{115.64^{\,3.90}+\ell^{\,3.90}},\quad R^2 = 0.9303.
\]
The steeper exponent ($n = 3.90$) produces a sigmoidal transition. MINDPixel operates most sensitively in the moderate-to-high illumination range. The two morphologies define complementary operating regimes for optical intensity inference.

\subsection*{Amplitude structure encodes spatial and temporal features under static illumination}

\begin{figure*}[t]
    \centering
    \includegraphics[trim={220 165 240 145},clip,width=\linewidth]{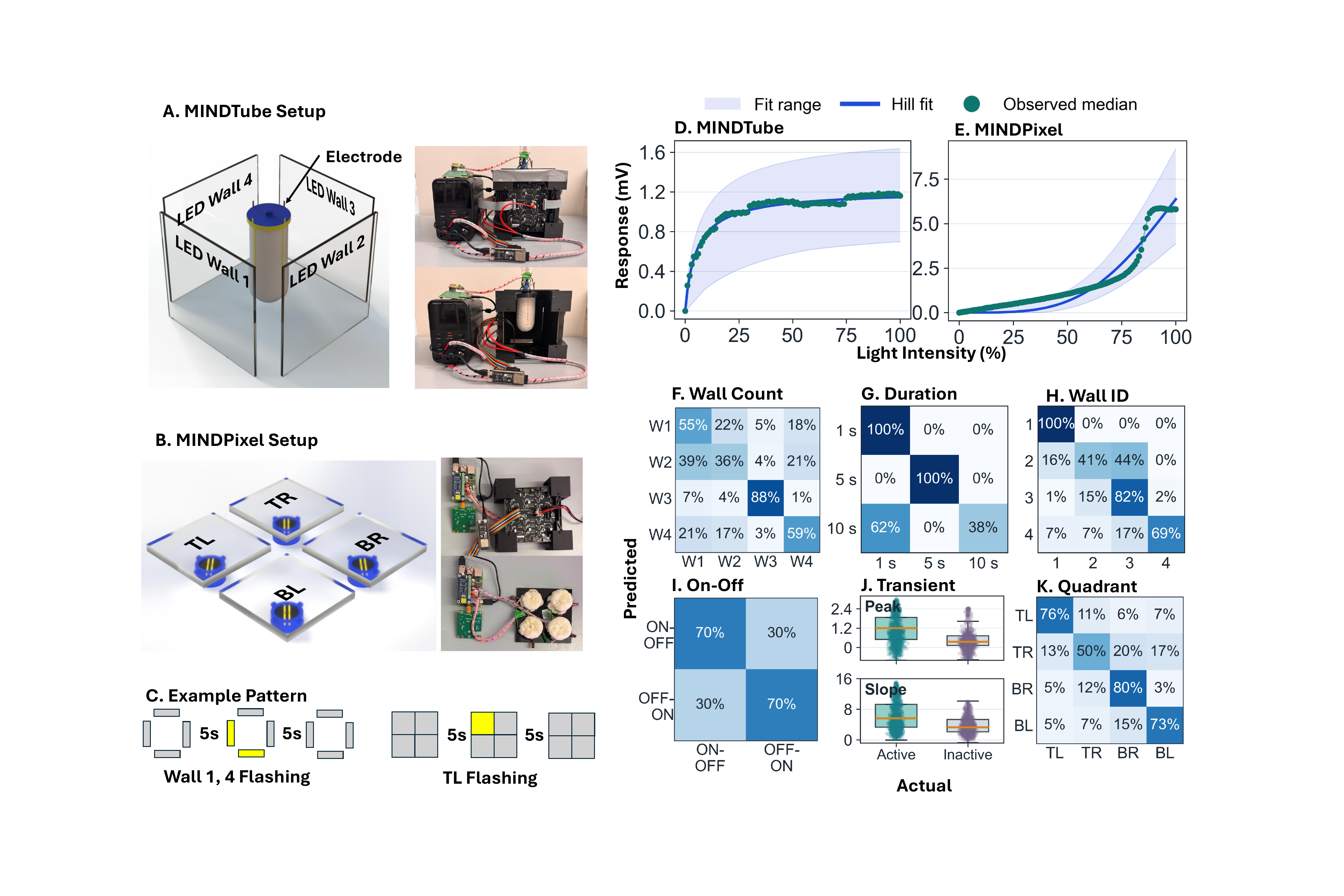}
    \caption{Static-light calibration and decoding in MINDTube and MINDPixel.
    (\textit{A}) MINDTube setup for four-wall illumination.
    (\textit{B}) MINDPixel setup for four-pixel illumination.
    (\textit{C}) Representative static-illumination protocols.
    (\textit{D}) MINDTube intensity--response calibration with observed medians and fitted Hill curve.
    (\textit{E}) MINDPixel intensity--response calibration with observed medians and fitted Hill curve.
    (\textit{F}) MINDTube wall-count confusion matrix, with 59.5\% accuracy.
    (\textit{G}) MINDTube duration-class confusion matrix, with 79.3\% accuracy.
    (\textit{H}) MINDTube wall-identity confusion matrix, with 73.2\% accuracy against a 25\% chance baseline.
    (\textit{I}) MINDPixel ON/OFF transition confusion matrix, with 70.0\% accuracy.
    (\textit{J}) MINDTube active-wall transient features exhibit larger scaled median peak responses (1.21 mV vs.\ 0.34 mV) and steeper scaled median initial slopes (5.67 mV vs.\ 3.32 mV).
    (\textit{K}) Four-quadrant confusion matrix across four MINDPixel units, with 69.8\% accuracy against a 25\% chance baseline.}
    \label{fig:static_setup_calibration_decoding}
\end{figure*}

With intensity calibrated, we asked what spatial and temporal information the same signal carries. The standardized four-channel circumferential geometry of MINDTube allows direct testing of this prediction across two flash paradigms. Local perturbations should produce distributed responses in which onset timing is shared across electrodes. Response amplitude should remain graded by proximity to the stimulus source. This predicts amplitude-based spatial encoding instead of a time-of-flight code. We tested this prediction across two flash paradigms (Fig.~\ref{fig:static_setup_calibration_decoding}C).

For MINDTube, one to four walls were independently illuminated for 1, 5, or 10\,s. Ridge regression on multichannel baseline-corrected voltage features decoded wall count with $59.5 \pm 6.8$\% accuracy and wall identity with $73.2 \pm 7.2$\% accuracy, compared with a 25\% chance baseline. Stimulus duration was identified at a balanced accuracy of $79.3 \pm 5.9$\% against a 33.3\% baseline. The experiments were conducted across three independent sessions with 21 experimental conditions, each repeated three times (Fig.~\ref{fig:static_setup_calibration_decoding}F--H; SI Appendix, Table~S1). The temporal profile of the bioelectrical signal encoded stimulus duration with high fidelity. 

Duration classification reached $79.3 \pm 5.9$\% balanced accuracy against a 33.3\% chance baseline (The duration confusion matrix (Fig.~\ref{fig:static_setup_calibration_decoding}G)). Because the duration classes were imbalanced, with 747, 29, and 42 trials for the 1, 5, and 10 s conditions, we also report overall accuracy and Cohen's $\kappa$. Overall accuracy reached 96.8\% with Cohen's $\kappa = 0.763$, indicating that performance was not explained solely by the dominant 1~s class.

Transient feature analysis confirmed the amplitude-encoding hypothesis directly. At the wall-event level, active-wall transients had larger scaled median peak responses (1.21 vs.\ 0.34) and steeper scaled median initial slopes (5.67 mV vs.\ 3.32 mV; peak: Wilcoxon $p = 0.031$, Cohen's $d_z = 1.16$; slope: $p = 0.031$, $d_z = 1.31$) than inactive-wall transients. Onset lag did not differ significantly at the session level (0.161 vs.\ 0.232~s; Wilcoxon $p = 0.313$). Therefore, the network produced a distributed response across all walls, with location encoded by channel-specific amplitude patterns rather than onset timing. A secondary, slower spatial modulation emerged after the initial transient. This component is consistent with Ca$^{2+}$ buffering dynamics evolving on timescales of seconds and carries additional location information beyond the initial per-channel amplitude differential~\cite{Itani2023LocalCalciumSignal}.

For MINDPixel, full-panel transition-direction classification reached $70.0 \pm 5.4$\% accuracy across 18 independent experiments (Fig.~\ref{fig:static_setup_calibration_decoding}I). Quadrant illumination across four spatially arranged MINDPixel units yielded $69.8 \pm 6.7$\% four-class accuracy against a 25\% chance baseline (Fig.~\ref{fig:static_setup_calibration_decoding}K). Therefore, MINDPixel encodes illumination transitions and retains decodable spatial structure.

\subsection*{Lagged decoding recovers continuous stimulus trajectory}

Static flashing showed that MIND encodes discrete spatial and temporal stimulus states. We next tested whether the same extracellular signal could track a continuously moving stimulus. This task requires the network response to preserve recent stimulus history while updating with changing stimulus position. Living networks possess finite adaptation and recovery timescales~\cite{Itani2023LocalCalciumSignal, Chen2010VIVIDWCC, Dasgupta2015PhotocycleLength}. Stimuli moving too slowly allow transient responses to decay before new spatial information arrives. Stimuli moving too quickly prevent coherent entrainment before the next cycle begins. Therefore, tested whether each morphology has an operational speed range in which stimulus motion and bioelectrical response dynamics remain aligned.

\begin{figure*}[t]
    \centering
    \includegraphics[trim={60 235 45 220},clip,width=\linewidth]{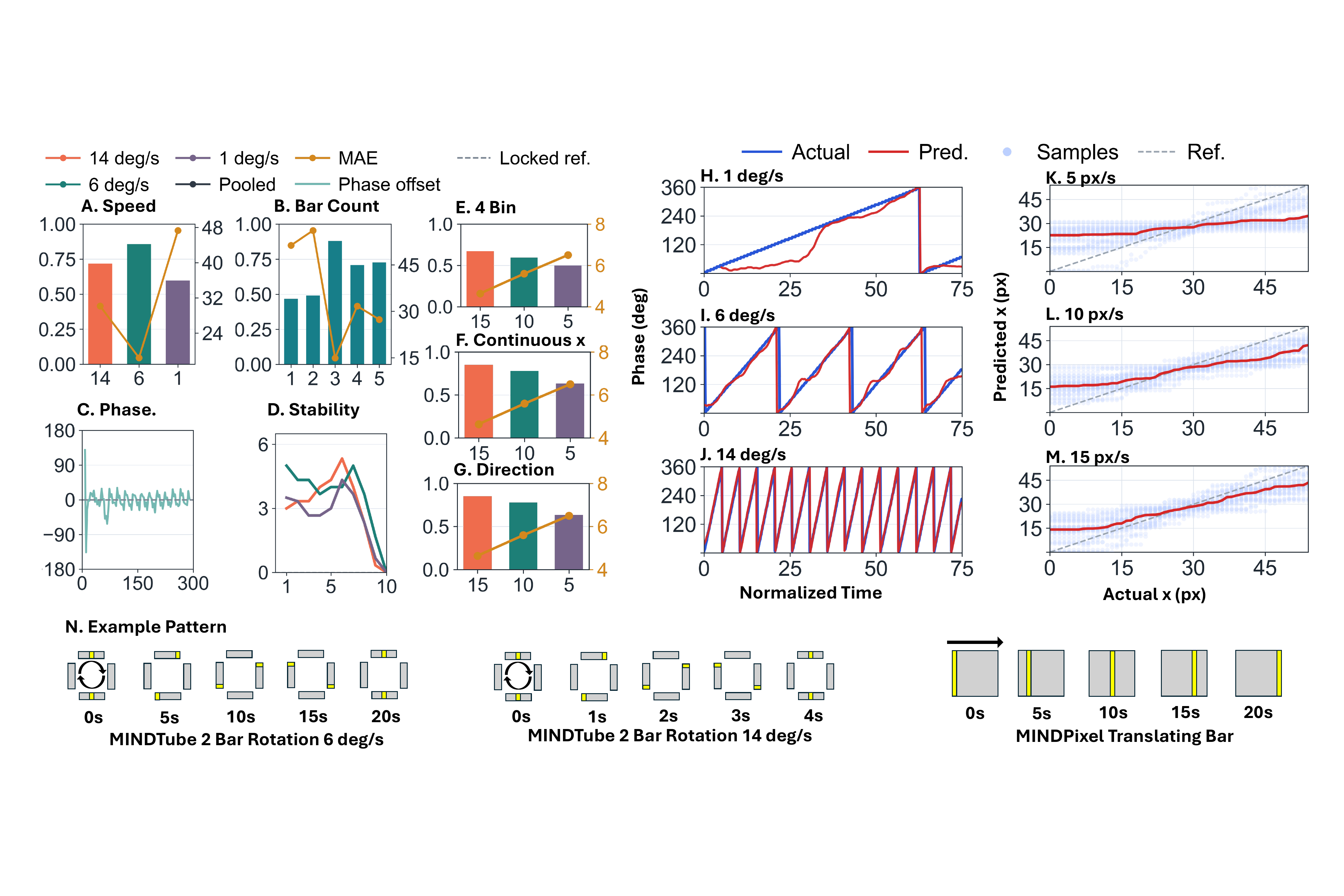}
    \caption{Continuous decoding of moving stimuli in MINDTube and MINDPixel.
    (\textit{A}) Lagged 4-bin decoding accuracy versus rotation speed for each MINDTube bar-count pattern.
    (\textit{B}) Decoding accuracy versus bar count at each speed.
    (\textit{C}) Representative reduced-phase trajectories with lagged decoder predictions for MINDTube.
    (\textit{D}) Phase-lock stability across conditions.
    (\textit{E}) 4-bin angular decoding accuracy by speed category.
    (\textit{F}) Continuous reduced-phase decoding performance by speed category.
    (\textit{G}) Direction classification accuracy versus rotation speed and bar count.
    (\textit{H--J}) Actual versus predicted reduced angular phase at 1, 6, and 14\,deg/s.
    (\textit{K--M}) Actual versus predicted translating-bar position at 5, 10, and 15\,px/s.
    (\textit{N}) Representative stimulus schedules for rotating-bar and translating-bar experiments.
    Across conditions, MINDTube achieved a median circular MAE of $22.15^\circ$, and MINDPixel achieved pooled position recovery with $R^2 = 0.7547$ and MAE of 11.16\,mm.}
    \label{fig:phase_lock}
\end{figure*}

Two moving-stimulus paradigms are summarized in Fig.~\ref{fig:phase_lock}. White-light bars rotated around MINDTube at 14, 6, and 1\,deg\,s$^{-1}$ across 33 independent sessions, with lagged ridge regression as the best-performing decoder. A vertical bar translated left to right across the MINDPixel panel at 5, 10, and 15\,px\,s$^{-1}$ across 27 independent sessions, decoded with the same lagged approach. In both cases, models incorporating recent signal history outperformed instantaneous decoders.

For MINDTube, angular position was analyzed in reduced phase $(n\rho) \bmod 2\pi$ for $n$-bar patterns (Fig.~\ref{fig:phase_lock}A). After excluding the first settling lap, matched-harmonic $R^2$ rose from $-0.0021$ to $0.3958$. Median lagged 4-bin accuracy reached $79.97 \pm 6.1$\% with median circular MAE of $22.15 \pm 6.84^\circ$. Three-bar patterns and medium-speed rotation at 6\,deg/s were most decodable, with full per-pattern and per-speed breakdowns in SI Appendix, Tables~S2 and S3. At the slowest speed tested (1\,deg/s), transients decayed before the next angular increment generated new decodable content, and entrainment quality directly predicted decoding performance across all tested conditions ($r = 0.8594$ with lagged 4-bin accuracy; $r = -0.8828$ with continuous phase error).

For MINDPixel, position recovery improved monotonically with speed, from the slowest tested condition to the fastest (Fig.~\ref{fig:phase_lock}). Pooled across all conditions, position $R^2$ was $0.7547 \pm 0.058$ with MAE of $11.16 \pm 1.42$\,mm. Discrete 4-bin accuracy improved from $36.8 \pm 6.2$\% without contextual history to $51.5 \pm 7.9$\% at eight history steps, indicating that the MIND-coupled fungal network retains positional state over sub-second timescales.

Across both morphologies, decoding was strongest when stimulus motion matched the response timescale of the living interface. Two temporal regimes underlie this behavior. Fast transients within 1 to 3\,s of any stimulus change encode spatial information through amplitude and slope differences. A slower sustained component was consistent with the intensity encoding captured by the Hill calibration. Together, these regimes show that fungal bioelectrical dynamics support inference about continuously evolving environments.

\subsection*{Co-occurring stimuli produce partial superposition rather than complete signal collapse}

\begin{figure*}[t!]
    \centering
    \includegraphics[trim={65 225 75 225},clip,width=.8\linewidth]{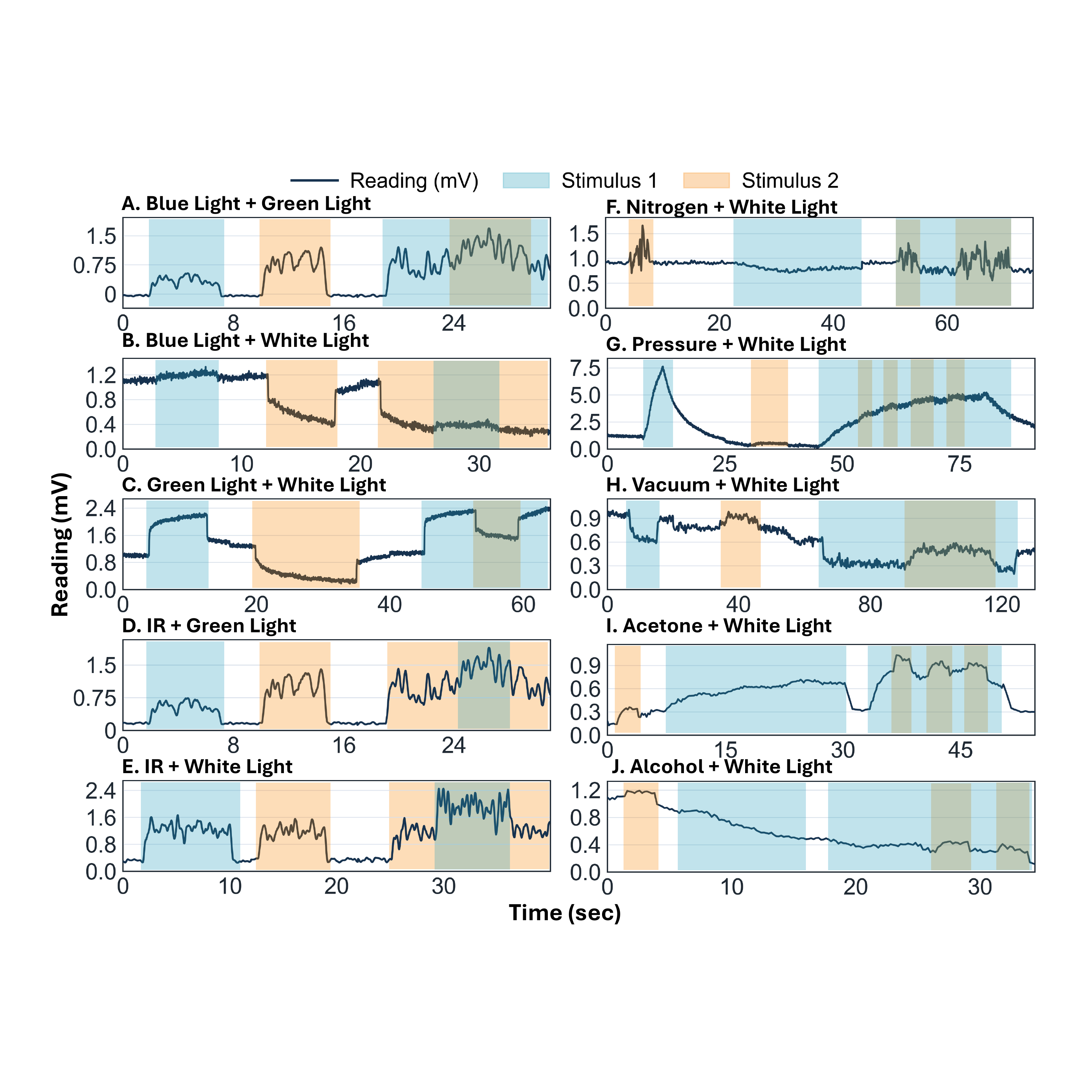}
    \caption{Representative responses to simultaneous stimuli.
    (\textit{A--J}) Example simultaneous-stimulus traces spanning optical, cross-modal, and chemical--optical pairs. Best joint-bit accuracy was 28.5\%, the highest within-pair accuracy was 71.4\%, and the best weighted single-stimulus template fit reached $R^2 = 0.738$.}
    \label{fig:combo}
\end{figure*}

Real-world deployment environments rarely present isolated stimuli. We therefore characterized MIND's response when two stimuli were applied concurrently. We analyzed 123 paired-stimulus events accumulated over eleven months of MIND operation. Representative responses appear in Fig.~\ref{fig:combo}.

Inverse Fast Fourier Transformation (IFFT)-derived features yielded the best joint-bit accuracy of $0.285 \pm 0.041$. The highest within-pair accuracy was $0.714 \pm 0.063$ for green light paired with white visible light, with full pair-level results in SI Appendix, Table~S5. Weighted single-stimulus template fits further confirmed the partial-superposition structure, reaching $R^2 = 0.738 \pm 0.082$ for the best-fitting pair.

Full-source separation from a merged channel is not supported by these data. The above-chance decoding indicates that co-occurring inputs are not collapsed into an indistinguishable composite waveform. With the given electrode geometry, simultaneous-stimulus responses are best described as partial superposition, with separable structure preserved alongside nonlinear coupling introduced by shared downstream ionic machinery~\cite{Zhao2024FungalPma1, Buffi2025ElectricalSignalingFungi}. Electrode geometry and channel selection influenced separability under simultaneous stimulation. These findings define a practical ceiling that denser electrode configurations may partially overcome.

\subsection*{Hill response structure generalizes across five phylogenetically diverse fungal species}

\begin{figure}[t]
    \centering
    \includegraphics[trim={1050 600 1050 575},clip,width=.85\linewidth]{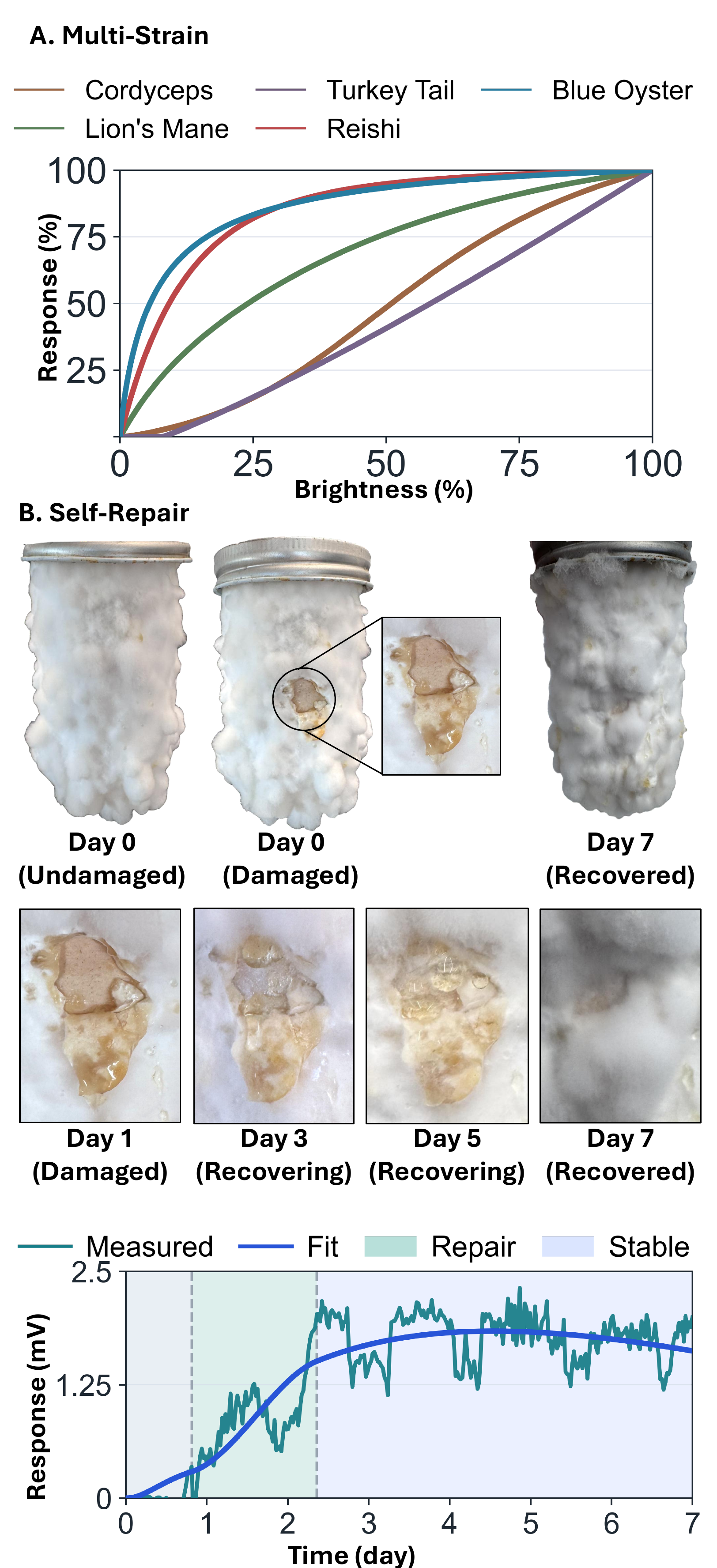}
    \caption{Cross-strain calibration and self-repair in MIND.
    (\textit{A}) Intensity--response calibration across five fungal strains grown on the same platform, with fitted Hill curves overlaid. All five strains produced consistent Hill fits, with inter-strain variation in apparent half-response brightness ($E$) and Hill exponent.
    (\textit{B}) Recovery after mechanical excision of a 20\,mm $\times$ 20\,mm active region, showing representative morphology and restoration of the evoked electrical response across three independent experiments.}
    \label{fig:repair_multistrain}
\end{figure}

All of the preceding results were obtained with blue oyster mycelium, so we next asked whether the calibrated response structure extends to other species. Core fungal electrophysiological machinery, including photoreceptors, Ca$^{2+}$ channels, and H$^+$-ATPase, is broadly conserved across fungal clades~\cite{Corrochano2019LightFungalWorld, Vu2015Cch1Mid1, Zhao2024FungalPma1}. This conservation predicts that a saturating Hill response should be shared across species grown on the same MIND interface. We cultivated four additional strains on the MINDTubes, including reishi (\textit{Ganoderma lingzhi}), cordyceps (\textit{Cordyceps militaris}), turkey tail (\textit{Trametes versicolor}), and lion's mane (\textit{Hericium erinaceus}), spanning distinct phylogenetic clades and ecological niches. All strains were evaluated at three months post-inoculation, each in triplicate across 15 total experiments (Fig.~\ref{fig:repair_multistrain}A; Table~\ref{tab:strain_hill}).

All five strains produced Hill-function fits with $R^2 = 0.89$--$0.93$ (Table~\ref{tab:strain_hill}). This shared saturating structure supports the interpretation that the calibration framework captures a conserved feature of fungal electrophysiology. The bioceramic platform supported consistent electrical coupling across all five strains, functioning as a universal interface for long-duration mycelial electrophysiology.

Inter-strain parameter variation confirmed that the platform resolves species-specific differences in calibrated sensitivity. Blue oyster and reishi produced the lowest apparent threshold values ($E = 6.75$ and $6.97$\%); turkey tail required the highest brightness before saturating ($E = 117.06$\%); cordyceps and lion's mane occupied an intermediate threshold but differed sharply in steepness ($n = 2.59$ vs.\ $0.95$). These parameter differences are consistent with strain-specific light-response biology reported across these taxa~\cite{Qi2020PoWC, GanodermaLingzhiWC2_2025, CordycepsCmWC1, CordycepsLightTranscriptome2023, HericiumErinaceusLight2025, TrametesVersicolorLight2025}. Because $E$ and $n$ are recoverable from calibration runs, strain selection becomes a practical design parameter for matching device sensitivity to a target illumination range.

\begin{table}[t]
\centering
\small
\caption{Hill-fit Parameters of Mycelial Clades on MINDTube}
\label{tab:strain_hill}
\begin{tabular}{lccc}
\toprule
Strain & $E$ (\%) & Hill $n$ & $R^2$ \\
\midrule
Blue Oyster & 6.75   & 0.90 & 0.93 \\
Reishi      & 6.97   & 1.10 & 0.93 \\
Lion's Mane & 61.09  & 0.95 & 0.92 \\
Cordyceps   & 62.29  & 2.59 & 0.91 \\
Turkey Tail & 117.06 & 1.73 & 0.89 \\
\bottomrule
\end{tabular}

\par\vspace{0.3em}

\begin{minipage}{\linewidth}
\footnotesize
\raggedright
$E$: Half-response brightness of the endpoint-normalized Hill fit (\% of full LED output).
$n$: Hill exponent. All strains were evaluated at three months post-inoculation ($n = 3$ per strain; 15 experiments total).
All experiments utilized MNMv3 mycoponic nutrient medium~\cite{Sanchez2026Mycoponics}.
\end{minipage}
\end{table}

\subsection*{Electrophysiological function recovers autonomously within 72 hours of mechanical damage}

Mechanical damage during field deployment requires the sensing substrate to recover without manual recolonization. MIND's continuous capillary nutrient supply through the bioceramic provides the metabolic support that regrowth requires. To test autonomous recovery, a 20\,mm $\times$ 20\,mm active mycelial region was mechanically excised across three independent MINDTube devices. Recovery was monitored over seven days (Fig.~\ref{fig:repair_multistrain}B). The ceramic chamber was filled with nutrient feed before excision to support regrowth. Recovery was assessed against three criteria: complete footprint re-coverage, restoration of baseline electrical activity, and maintained stimulus responsiveness. Hyphal bridging of the wound margins commenced within 24 hours. By Day 3, continuous surface coverage was re-established, and stimulus responsiveness was fully restored. By Day 7, the repaired region was morphologically indistinguishable from undamaged areas. The plotted recovery trajectory represents the mean across all three devices.

Electrical recovery preceded full visible restoration by approximately two days. Electrical activity began recovering by Day 1. The fitted recovery curve reached 95\% of the Day-7 plateau by Day~3, consistent with anastomosis restoring ionic continuity before full surface recolonization~\cite{Buffi2025ElectricalSignalingFungi,Fukasawa2024WeekLongOscillation}. These results indicate that functional recovery depends on restoration of a connected living pathway, rather than complete recovery of surface density. Continuous nutrient supply through the porous bioceramic supported this regrowth.


\section*{Discussion}
MIND demonstrates that an engineered substrate-electrode platform is the enabling condition for converting living mycelium into a calibrated multimodal sensor. The porous bioceramic with gravity-fed mycoponic perfusion sustains mycelial metabolism without substrate replacement, while the passive Cu-Kapton differential electrode array standardizes extracellular coupling across months of colonization. A modular geometry implements the same bioceramic-electrode principle in two physical forms, cylindrical (MINDTube) and planar (MINDPixel), and this two-morphology design confirms that the engineering solution generalizes across substrate geometries. Together, both morphologies sustained calibration-quality recordings for more than eleven months. A single unmodified device distinguished fourteen stimulus classes spanning five sensory domains. Hill-type intensity calibration reproduced across five phylogenetically diverse fungal clades on the identical interface. Multichannel decoding recovered stimulus identity, location, duration, and continuous trajectory from bioelectrical output alone. The same perfusion architecture that maintains long-term viability restored full electrophysiological function within seventy-two hours of mechanical excision.

Nutrients flow through 0.3 $\mu m$ pores to the mycelial surface while the pore geometry mechanically excludes most contaminants~\cite{Sanchez2026Mycoponics}. This eliminates dehydration, metabolite depletion, and contamination that limit Petri-dish preparations to days or weeks~\cite{Adamatzky2022FungalElectronics, Buffi2025ExternalStimuli}. The same continuous supply sustained regrowth across the 20 mm × 20 mm excision wound. Electrical recovery preceded full surface restoration by approximately two days, consistent with anastomosis re-establishing ionic continuity before complete morphological repair (~\cite{Buffi2025ElectricalSignalingFungi, Fukasawa2024WeekLongOscillation}. Functional recovery on this timescale is unreachable without nutrient delivery through the damage event itself. The Cu/Kapton electrode architecture allows the colonizing mycelium to grow while Kapton insulates the full trace length. This three-layer protection scheme supported calibration-quality readout across the full eleven-month window without electrode replacement. The four-channel circumferential geometry of MINDTube and the four-pixel arrangement of MINDPixel make the amplitude-based spatial code measurable. A single-channel recording would collapse the distributed spatial structure that multichannel ridge regression recovers, and the 400 SPS acquisition rate is sufficient because the spatial code is amplitude-based rather than latency-based.

The main limitations of the current implementation arise from its electrode design and operating conditions. Four differential channels per MINDTube and four discrete pixels per MINDPixel were sufficient to demonstrate multimodal decoding and spatial inference in these proof-of-concept architectures, but the mycelial network distributes stimulus-evoked amplitude gradients at finer spatial scales than this channel count can resolve. Increasing electrode density, while balancing nutrient delivery to the mycelium layers, would improve spatial resolution and simultaneous stimulation separability in parallel. Also, the baseline drift following repeated stimulation is small relative to the smallest evoked deflections analyzed in our characterization. Such small shifts are routinely handled by adaptive baseline tracking. Periodic recalibration may be required for long-duration deployments, especially under field conditions. The 72-hour self-repair window represents a genuine biological self-maintenance capability that scales with nutrient availability and the severity of damage. Its practical relevance depends on whether the target application can tolerate a repair interval of that timescale, which varies widely across deployment scenarios.

Overall, the results motivate two parallel tracks of future work. On the engineering side, higher-density electrode arrays and tiled MINDPixel configurations represent a path to improved spatial resolution and simultaneous-stimulus separability, as mycelium already encodes stimulus information at finer spatial scales than the current four-channel geometry can sample. On the biological side, strain selection represents an additional design dimension that can be matched to target modalities and sensitivity requirements before fabrication, with the EC$_50$ and Hill-exponent variation reported here as a practical tuning parameter. Beyond these performance dimensions, the primary sensing element of MIND is a living fungal network composed of biodegradable biopolymers cultivated on a kiln-fired ceramic substrate. Whether these material properties translate into a quantifiable lifecycle advantage over conventional solid-state sensor suites covering equivalent stimulus domains remains an open question that a full comparative assessment, accounting for nutrient medium production, ceramic firing, and operational power draw alongside fabrication inputs, would need to address.

The broader implication is that MIND is the interface that converts the organism into an instrument. MIND demonstrates that a single platform built around mycoponic perfusion, passivated differential electrodes, and a modular substrate geometry yields multimodality, calibration stability, cross-strain generalizability, and autonomous repair as joint properties of one device. This positions fungal electrophysiology as a calibrated measurement platform for sensing applications in which the stimulus set, the electrode layout, and the failure modes cannot be fully specified in advance.

\matmethods{

\subsection*{Mycoponic Bioceramic Platform}

The bioceramic substrate was engineered to act as a mechanical filter that is anti-microbial but porous to support long-term mycelial viability~\cite{Sanchez2026Mycoponics}. A powder blend (wt.\,\%: 30 china clay, 30 ball clay, 15 CaCO\textsubscript{3}, 25 diatomaceous earth) was sieved, dispersed in water with a polyacrylate dispersant and a cellulose-ether binder, and vacuum de-aired before forming. For the MINDTube morphology, the slurry was pressure-extruded to $L = 90$~mm, $OD = 50$~mm, wall thickness $t = 7$~mm, and fired at 1100$^\circ$C. The fired body exhibited a median pore size of approximately 0.3~$\mu$m, providing mechanical exclusion of most bacteria. Fluid-port end pieces were press-fitted to each end face for gravity-fed nutrient delivery. For the MINDPixel morphology, the substrate was formed as a flat ceramic disk of 30~mm radius and 4~mm thickness. Key geometry and electrode dimensions for both morphologies are summarized in Fig.~\ref{fig:overview_workflow}.

\subsection*{Electrode Fabrication and Integration}
Electrodes were fabricated as a copper-Kapton laminate. Copper tape was uniformly spray-inked and laser-scribed into 80~mm $\times$ 1.5~mm rectangular traces, then processed by standard wet chemical etching. Kapton insulation tape was precision-cut using a CNC machine. The laminated electrodes were transferred onto the bioceramic surface using a positioning guide, with Kapton covering the full trace length except for a 1.5~mm $\times$ 1.5~mm exposed copper contact pad at the sensing interface.

For MINDTube, four electrode pairs were placed equidistantly around the outer circumference, positioned 70~mm axially along the lower portion of the tube where fruiting bodies concentrate; each pair formed one differential recording channel. For MINDPixel, each pixel carried one localized differential interface, so a four-pixel arrangement yielded four simultaneously recorded channels. Once colonized, the mycelium grows over the contact pad. No electrode degradation was observed across the full eleven-month operational window.

\subsection*{Inoculation and Cultivation}

Assembled bioceramic modules were sterilized by autoclaving and inoculated with mycelial liquid culture of blue oyster mycelium (\textit{Pleurotus ostreatus}) as the primary experimental strain~\cite{Sanchez2026Mycoponics}. Similarly, other strains were also innoculated onto the autoclaved tubes and fed with MNM v3, with growth proceeding for approximately two weeks under low light ($<50$ lux), high moisture (70--90\% relative humidity), and cool ambient laboratory conditions (18--22$^\circ$C) until stable colonization was established. During operation, a custom nutrient blend based on Mycoponic Nutrient Medium V2 was gravity-fed via the fluid-port end piece (MINDTube) or through a sealed chamber reservoir (MINDPixel), from which nutrients diffused through the porous ceramic to sustain the mycelial layer. For multi-strain experiments, four additional species were cultivated on the same platform: reishi (\textit{Ganoderma lingzhi}), cordyceps (\textit{Cordyceps militaris}), turkey tail (\textit{Trametes versicolor}), and lion's mane (\textit{Hericium erinaceus}). Additional details on the mycoponics materials and protocols are described in Porterfield et al.~\cite{Sanchez2026Mycoponics}.

\subsection*{Signal Acquisition and Stimulus Delivery}

Bioelectrical signals were acquired with an ADS1263-based 32-bit delta-sigma conversion chain operated in differential mode at 400~SPS per channel using the converter's internal 2.5~V reference. Stimulus control and electrophysiological acquisition shared the same Raspberry Pi controller and software time base, ensuring precise alignment between stimulus delivery and voltage recording. MINDTube experiments used four simultaneous differential channels; MINDPixel experiments used one channel per pixel, with four channels recorded simultaneously for full four-pixel configurations.

A 20~s baseline segment was recorded at the start of each session. Local baseline subtraction and moving-mean smoothing were applied before feature extraction and decoding. Detailed preprocessing parameters, circuit diagrams, and source-control scripts are provided in the \textit{SI Appendix}.

Optical stimuli were delivered as electromagnetic radiation using a programmable 64\,$\times$\,64 RGB LED matrix for the blue-light ($\lambda \approx 470$\,nm), green-light ($\lambda \approx 525$\,nm), and white visible-light (approximately 400--700\,nm) conditions, together with auxiliary near-infrared (IR; $\lambda \approx 850$\,nm), ultraviolet B (UVB; $\lambda \approx 311$\,nm), and ultraviolet C (UVC; $\lambda \approx 254$\,nm) sources. For MINDTube, four LED panels arranged around the tube circumference enabled independent wall illumination and continuous rotation of light bars. For MINDPixel, a single panel was positioned directly above the disk. Programmed LED output percentage served as the ground-truth intensity reference for all calibration experiments, with 0\% corresponding to LED off and 100\% to full programmed panel output. Ambient visible-light level was monitored with a VEML7700 ambient-light sensor; visible and IR optical output was verified with a TSL2591 light-to-digital sensor. UVB source output was verified with a VEML6075 UV sensor, and UVC source output was verified with a filtered 254~nm photodiode (S12742-254).

The stimulus bank encompassed chemical stimuli (acetone, isopropyl alcohol, carbon dioxide (CO\textsubscript{2}), propane, MAP-Pro gas (primarily propylene/propene, C\textsubscript{3}H\textsubscript{6})), optical stimuli (white visible light, IR, UVB, and UVC), mechanical stimuli (touch and pressure loading), thermal stimuli (temperature), and biological perturbations (tar spot of maize caused by \textit{Phyllachora maydis}). Detailed source specifications, chamber geometries, and timing schedules are provided in the \textit{SI Appendix}.

For stimuli admitting controlled delivery, chamber conditions were monitored continuously with environmental and stimulus-specific transducers. Pressure, vacuum, chamber temperature, and relative humidity were monitored with a BME680 environmental sensor. CO\textsubscript{2} concentration was monitored with an SCD41 CO\textsubscript{2} sensor. Propane-rich gas exposure was monitored with a TGS2610-D00 LP-gas sensor; the same sensor served as a chamber-exposure proxy during MAP-Pro welding gas trials. Delivery rate was set with a mass-flow-controlled feed and verified against the commanded setpoint throughout each application window. Within each rate group, the instantaneous rate of change was derived from the corresponding sensor time series and used to construct response envelopes for pressure, vacuum, and gas-exposure sweeps. Environmental variables not under direct manipulation, including ambient room temperature (20--24$^\circ$C), relative humidity (40--60\%), and background light level (${<}50$ lux during optical sessions; 200--400 lux during non-optical sessions), were monitored throughout and maintained within these bounds to ensure that observed bioelectrical responses reflected the intended stimulus class.

\subsection*{Experimental Units and Replication}

A \textit{device} denotes one living MINDTube or one living MINDPixel unit. A \textit{session} denotes one continuous recording run on one device under one defined experimental protocol. An \textit{event} (or \textit{trial}) denotes one stimulus presentation within a session. Unless otherwise stated, experimental support is reported in terms of independent devices, independent sessions on those devices, or event-level counts pooled across sessions. Static-light and moving-light decoding experiments drew repeated stimulus events from the same living device across multiple sessions. Multi-strain experiments were conducted on biologically distinct devices prepared from different fungal strains. Self-repair measurements were performed on three independent MINDTube devices, and the plotted recovery trajectory represents the average across all three.

\subsection*{Self-Repair Protocol}

A $20\,\text{mm}\times20\,\text{mm}$ region of the active mycelial surface was mechanically excised and the ceramic chamber was filled with nutrient feed prior to excision to support regrowth. Recovery was monitored over seven days by flashing a concentrated light beam for 5~s onto the damaged area every 30~minutes and recording the maximum response amplitude. Recovery was assessed against three criteria: complete re-coverage by visual inspection, restoration of baseline electrical activity, and maintained responsiveness to stimulation.

\subsection*{Tar Spot of Maize Exposure}

A sealed chamber was sprayed with isopropyl alcohol, then hydrogen peroxide, and placed in a UVC sterilization chamber for 24~hours. Harvested maize leaves (\textit{Zea mays}) carrying tar spot lesions caused by \textit{Phyllachora maydis} were then placed inside the chamber containing an active MIND device, and the bioelectrical response was recorded under the standard acquisition protocol.

}

\showmatmethods{}

\dataavail{Code and data are available upon request.}

\acknow{
We thank Simone X. Moulton, Adriana K. Sanchez, Tayla Koenig, and Richard J. Barker for assistance with biological preparation, culture handling, and preparation of mycoponic nutrient media. 
We thank Hoa Duc Tri Nguyen, Minh Hoang Giang Nguyen, Kailey Christina Dvorak, and Jonas Carl Erickson for assistance with stimulus-delivery trials and data-collection support. 
This work received no external funding.
}

\showacknow{}

\bibsplit[15]

\bibliography{references}

\end{document}